# The Stromlo-APM Redshift Survey II.
## Variation of Galaxy Clustering with Morphology and Luminosity

**J. Loveday**
Fermi National Accelerator Laboratory, PO Box 500, Batavia, IL 60510, USA
loveday@fnal.gov

**S.J. Maddox**
Royal Greenwich Observatory, Madingley Road, Cambridge, CB3 0EZ, England
sjm@mail.ast.cam.ac.uk

**G. Efstathiou**
Department of Physics, Keble Road, Oxford, OX1 3RH, England
gpe@oxds02.astro.ox.ac.uk

**B.A. Peterson**
Mount Stromlo and Siding Spring Observatories, Weston Creek PO, ACT 2611, Australia
peterson@mso.anu.edu.au

# Abstract

We present clustering measurements for samples of galaxies selected by morphological type and luminosity from the recently completed Stromlo-APM Redshift Survey. We find very different results between real and redshift-space estimates of the correlation function. The real space correlation function for the all-galaxies sample is well fit on scales $0.2$–$20 h^{-1}$Mpc by a power-law with slope $\gamma_r = 1.71$ and correlation length $r_0 = 5.1 h^{-1}$Mpc. In redshift space the slope is shallower, $\gamma_s = 1.47$ and the correlation length is slightly higher, $s_0 = 5.9 h^{-1}$Mpc.

Early type galaxies are clustered more strongly by a factor 3.5–5.5, than late type galaxies. In real space the slope of the correlation function for early type galaxies is $\gamma_r = 1.85$, slightly steeper than for late types, $\gamma_r = 1.64$. In redshift space however, early type galaxies have a very shallow correlation function slope, $\gamma_s = 1.25$. This implies that these early-type galaxies suffer from enhanced redshift-space distortions compared to late-type galaxies.

Low-luminosity galaxies are clustered more weakly by a factor of $\sim 2$ than $L^*$ and brighter galaxies on scales $\gtrsim 1 h^{-1}$Mpc. Also the slope of the correlation function is steeper for low-luminosity galaxies, so that the amplitude is a factor 4 lower at $10 h^{-1}$Mpc. No difference, however, is seen between the clustering of $L^*$ and more luminous galaxies, an observation which may be hard to reconcile with some theories of biased galaxy formation. Both redshift-space and real-space clustering estimates show a similar dependence on luminosity.

Our results hint that luminosity segregation may be a purely primordial effect, due to a lower bias factor for low-luminosity galaxies, whereas morphological segregation, being most apparent on scales $\lesssim 1 h^{-1}$Mpc, may be enhanced by environmental factors.

Subject headings: galaxies: clustering — galaxies: fundamental parameters — large-scale structure of Universe

# 1  Introduction

The dependence of galaxy clustering on luminosity and morphology is of great relevance to all galaxy formation theories and in understanding the large-scale matter distribution in the universe. If there are differences between the clustering of various types of galaxies we can immediately infer that at least one of the galaxy types is a biased tracer of the underlying mass distribution. In general we expect all galaxy samples to be biased at some level relative to the mass, and the differences in clustering can test various models for the bias of galaxies relative to mass. For example, the process of "natural bias" (White *et al.* 1987) leads to a galaxy correlation function that is a constant factor times the mass correlation function, with the factor being larger for more massive galaxies. More realistically, galaxy formation depends on complex processes which involve the local environment, some feedback mechanisms or galaxy interactions, as well as the depth of the dark matter potential (*e.g.* Dekel and Rees 1987). Quantitative measurements of the relative distribution of galaxies of different luminosities and types will tightly constrain models of these processes.

That the correlation function for elliptical and lenticular galaxies has a steeper slope and larger amplitude on small scales than the correlation function for spiral galaxies has been known at least since Davis and Geller (1976) calculated the angular correlation function of morphologically-selected galaxy subsamples from the Uppsala Catalogue (Nilson 1973). The possible dependence of galaxy clustering with luminosity has been much more controversial. Several groups (*e.g.* Bothun *et al.* 1986, Binggeli *et al.* 1990, Eder *et al.* 1989, Thuan *et al.* 1991, Weinberg *et al.* 1991) have found that the clustering of dwarf galaxies is consistent with that of bright galaxies (or at least that dwarf galaxies do not 'fill the voids' in the bright galaxy distribution). Alimi *et al.* (1988) and Phillipps and Shanks (1987) have measured clustering strength over a range of luminosities and find no evidence for enhanced clustering of more luminous galaxies. On the other hand, Davis *et al.* (1988), Hamilton (1988), Salzer *et al.* (1990), Santiago and da Costa (1990), Iovino *et al.* (1993) and Park *et al.* 1994 *do* claim to detect stronger clustering of bright galaxies compared with faint. Maurogordato and Lachieze-Rey (1991) find luminosity-dependence in the void probability function but not in the two-point correlation function. Hasegawa and Umemura (1993), after extinction-correcting CfA data, find weak luminosity segregation of opposite sign in early and late type galaxies.

All of the above analyses use fairly shallow ($m_{\rm lim} \sim 15$) catalogues of galaxies so that intrinsically faint galaxies can only be seen within a very small, nearby volume. It has thus been difficult to make a reliable comparison between clustering properties of bright and faint galaxies. In this paper we analyse the Stromlo-APM Redshift Survey which samples a much larger volume of space than any previous optically-selected galaxy redshift surveys; $L^*$ galaxies can be seen out to a distance $\approx 180h^{-1}{\rm Mpc}$, that is within a volume $\sim 2.5$ million $h^{-3}{\rm Mpc}^3$. This large volume has been surveyed rapidly by using a 1 in 20 sparse sampling strategy to select galaxies for redshift measurement. The uniform sampling of such a large volume makes the Stromlo-APM Survey an extremely powerful sample for studying luminosity segregation in galaxy clustering.

The construction of the survey has been briefly described in an earlier paper (Loveday *et al.* 1992a, hereafter Paper 1) and will be described in full in a future paper in this series. An analysis of the large-scale clustering of galaxies in the Stromlo-APM Redshift Survey and a comparison with two versions of the Cold Dark Matter theory has been given by Loveday *et al.* 1992b (hereafter LEPM). In the present paper we study the clustering properties of subsamples of galaxies selected from the survey by morphological type and luminosity. The paper is organised as follows. In section 2 we describe the galaxy samples used in the clustering analyses. In section 3 we discuss two estimators for the redshift-space correlation function $\xi(s)$, one dependent on and one independent of an assumed mean galaxy density and apply these estimators to the galaxy samples listed in section 2. In section 4 we present two estimates of the real-space correlation function $\xi(r)$ unaffected by redshift-space distortions. Our conclusions are presented in section 5. Throughout the paper, we use $r$ to denote real-space separations and $s$ to denote separations in redshift-space. Unless otherwise stated, error bars in figures and quoted errors in numerical quantities are 1-sigma dispersions calculated by analysing nine bootstrap-resampled versions of the survey (Barrow *et al.* 1984).

## 2 The Galaxy Samples

The Stromolo-APM Redshift Survey consists of 1787 galaxies with $b_J \leq 17.15$ selected randomly at a rate of 1 in 20 from the APM Galaxy Survey (Maddox *et al.* 1990a,b). The survey covers a solid angle of 1.3 Sr (4300 square degrees) in the south galactic cap. The APM magnitudes have been calibrated and corrected for photographic saturation using CCD photometry as described in Paper 1. An approximate morphological type was assigned to each galaxy by visually inspecting the images on the UKST survey plates. Redshifts have been obtained with the MSSSO 2.3m telescope at Siding Spring. Measured radial velocities are transformed to the local group frame using $v = v + 300 \sin(l) \cos(b)$ and we assume $\Lambda = 0$, $q_0 = 0.5$ and $H_0 = 100$ km s$^{-1}$Mpc$^{-1}$ with uniform Hubble flow in calculating distances and absolute magnitudes. We adopt $k$-corrections for different morphological types in the $b_J$ system as described by Efstathiou, Ellis and Peterson (1988). More details about the survey are given in Paper 1, and the construction will be described in full in a future paper in this series.

We draw six samples from the Stromlo-APM Redshift Survey: (a) all galaxies; (b) low, (c) medium and (d) high luminosity galaxies; (e) early and (f) late morphological types. These samples are defined in Table 1 and in Fig. 1 we show the redshift-distance histogram for each sample. Note that the volumes of each sub-sample all overlap to some extent, and that even the faintest sample extends beyond $100h^{-1}$Mpc in depth. For the luminosity-selected samples (b, c and d) we have applied an apparent magnitude bright limit, $m \geq 15$, since galaxies brighter than 15th mag suffer from photographic saturation on deep Schmidt plates and hence unreliable magnitudes. The luminosity limits were chosen to divide the galaxies into sub-$L^*$, $\sim L^*$ and super-$L^*$ samples with roughly equal numbers per sample.

## 3 The Redshift Space Correlation Function $\xi(s)$

### 3.1 Estimators

In any flux-limited survey, the observed density of galaxies will decrease with distance $x$ from the observer. In order that an estimate of the correlation function not be dominated by the nearby galaxies, it is important to give the appropriate weight to each galaxy. The variance in the estimated $\xi(s)$ is minimised by weighting each galaxy in a pair at redshift-space separation $s$ as

$$w_i = \frac{1}{[1 + 4\pi J_3(s) n(x_i)]}, \quad J_3(s) = \int_0^s s^2 \xi(s) ds, \qquad (1)$$

(See Appendix; this weighting scheme was first used by Efstathiou (1988)), where $n(x_i)$ is the mean galaxy density at the distance $x_i$ of the i'th galaxy. We determine $n(x_i)$ by integrating our estimate of the observed luminosity function, allowing for the effects of magnitude errors as described in Paper 1 (eq. [3]). Histograms of these predicted distributions are plotted as the dotted lines in Fig. 1.

To apply this weighting scheme, we need a model for $J_3(s)$. As we showed in LEPM, large-scale clustering in the Stromlo-APM survey, and in other surveys, is well described by the linear power-spectrum of an $\Omega h = 0.2$ scale-invariant CDM model, hence we have calculated weights from (1) using this model. We find almost identical estimates of $\xi(s)$ if $J_3$ for an $\Omega h = 0.5$ CDM model is used to calculate the weights. The estimates of $\xi(s)$ on large scales are not sensitive to the weighting scheme provided realistic values for $J_3$ are used.

In order to allow for the survey boundaries and selection function, we generate a random catalogue which fills the same volume as the galaxies and has the same selection function. Random points are generated within the distance range $x_{min}$ to $x_{max}$ according to a selection function obtained by integrating the observed luminosity function (Paper 1, eq. [11] and Fig. 4b). We choose $x_{min} = 5h^{-1}$Mpc, $x_{max} = 400h^{-1}$Mpc and a ratio of random points to galaxies $N_r/N_g \approx 10$. The sky coordinates of the points are chosen from a uniform distribution over each survey field.

The standard estimator for $\xi(s)$ is

$$1 + \xi(s) = \frac{n_r}{n_g}\frac{w_{gg}(s)}{w_{gr}(s)}, \qquad (2)$$

where $w_{gg}(s)$ is the summed product of the weights (1) of galaxy pairs in separation bin $s$, $w_{gr}(s)$ is the equivalent quantity for galaxy-random pairs and $n_g$ and $n_r$ are the mean densities of galaxies and random points, calculated with a minimum-variance estimator (Paper 1, eq. 4). The problem with this estimator is how one copes with fluctuations in the galaxy density $n_g$. By using a large enough random sample, fluctuations in $n_r$ can be made negligible, but for real redshift surveys, the actual density $n_g$ for a subsample may not correspond to the expected density $n_e$ for a homogeneous Universe. See, for example, Davis *et al.* (1988) and Maurogordato and Lachieze-Rey (1991) for a discussion of this problem.

Hamilton (1993) has pointed out that one can measure $\xi(s)$ independently of any assumed galaxy density with the following estimator

$$1 + \xi(s) = \frac{w_{gg}(s)w_{rr}(s)}{[w_{gr}(s)]^2}. \qquad (3)$$

Here $w_{rr}(s)$ is the summed product of the weights of random-random pairs. Note that the relative densities of galaxy and random points measured at separation $s$ are automatically accounted for by this estimator — there is no need to assume an overall galaxy density $n_g$. In this respect this estimator is similar to the 'ensemble' estimator of the angular correlation function $w(\theta)$ measured from counts of galaxies in cells

$$1 + w(\theta) = \frac{\langle N_i N_j \rangle}{\langle N_i \rangle \langle N_j \rangle}. \qquad (4)$$

These estimators are affected only to second order by density fluctuations related to the sample boundaries, whereas 'direct' estimators are affected to first order by such density fluctuations.

Of course, when comparing $\xi$ between subsamples of a catalogue, one must be wary of the effects of such fluctuations in galaxy density because different parts of space are probed by the different samples. For example Fig. 1 shows that the overlap in volume between samples (b) and (d) is rather small, and that most of the galaxies are from independent volumes. Table 1 shows that the actual galaxy density $n_g$ varies by up to 15% from the expected mean density $n_e$ (given simply by integrating the luminosity function over the appropriate magnitude range) for the luminosity-selected samples, and so the observed behaviour of $\xi(s)$ determined with (2) at small amplitudes must be interpreted with some caution. One might choose to use the expected density $n_e$ in equation (2) rather than the actual density $n_g$ when normalising $1 + \xi$. However, this procedure leads to a positive 'tail' in $\xi$ for over-dense samples and a negative 'tail' for under-dense samples, thus making comparison between samples difficult. By using the density-independent estimator (3), one does not have to assume a density for calculating $\xi(s)$, normalisation is determined automatically from those galaxies at each given separation. Hence the density-independent estimates should be much more reliable than the density-dependent estimates.

In Fig. 2 we plot $\xi(s)$ measured from the Stromlo-APM survey using both density-dependent (2; open symbols) and density-independent (3; solid symbols) estimators. In Fig. 2a we see that the density-dependent estimator finds considerably more power on large scales than the density-independent estimator, even when applied to the whole survey for which $n_g$ is relatively well determined. The difference in estimates is due to the slight mis-match between the radial density functions of the galaxies and the random catalogue (Fig. 1a) — there is a slight underdensity in galaxies on scales $\gtrsim 200h^{-1}$Mpc compared with what we would expect from the best-fit Schechter luminosity function. This slight mis-match is detected by the density-dependent estimator as increased large-scale clustering. The density-independent estimator is much less sensitive to large-scale gradients in the data, in this case large-scale gradients in the relative galaxy/random density.

A possible concern with the density-independent estimator is that it is removing intrinsic large-scale power in the galaxy distribution, not just artificial gradients due to uncertainty in the selection function. We can address this concern by analysing a volume-limited subsample of the catalogue, since one does not need to know the selection function to analyse a volume-limited (i.e. uniform density) sample. Additionally, no variable-weighting scheme is necessary for such a sample. Fig. 3 shows the redshift-space two-point correlation function measured from a sample volume-limited to $200h^{-1}$Mpc (428 galaxies brighter than $M_{b_J} = -19.7$). We now see very little difference between the density-dependent and density-independent estimators, suggesting that the density-independent estimator is not seriously 'filtering out' large-scale power. In fact, the density-independent estimator shows very slightly *enhanced* large-scale power over the density-dependent estimator for this sample.

While the volume-limited sample shows slightly more large-scale power than the density-independent estimate from the full sample, it does not show the very large-scale clustering at $s \approx 100h^{-1}$Mpc apparently detected by the density-dependent estimate from the full sample. Note the density-dependent estimate (Fig. 2a, open symbols) shows more large-scale power than our earlier determination of $\xi(s)$ (LEPM, Fig. 3). For this earlier analysis, only galaxies within $300h^{-1}$Mpc distance were used, and the selection function was obtained by integrating the best-fit pure Schechter luminosity function. As discussed in Paper 1, a Schechter function convolved with a Gaussian helps correct for random magnitude errors in the data and provides a better fit to the observed luminosity function. It is the combined effect of a slightly different selection function and including galaxies beyond $300h^{-1}$Mpc which yields the increased large-scale power found with the density-dependent estimator in Fig. 2a. The density-independent estimator in Fig. 2a is in good agreement with the earlier (density-dependent) estimate in LEPM (which barely changes if a density-independent esti-mator is used). The lack of sensitivity to the limiting distance and exact form of selection function is an important advantage of the density-independent estimator over the density-dependent one.

## 3.2 Comparisons Between Galaxy Samples

The clustering measurements using both density-dependent and density-independent estimators for all six galaxy samples listed in Table 1 are presented in Figure 2. The dotted lines in Fig. 2 show $\xi(s)$ predicted by $\Gamma = 0.2$ biased CDM linear theory to aid comparing the samples.

The density-dependent estimates show large variations on scales $s \gtrsim 20h^{-1}\text{Mpc}$, with a trend of increasing large-scale power with luminosity. As discussed in the previous section, the density-dependent estimates are sensitive to errors in the estimated galaxy density $n_g$ and we believe this apparent strong trend is caused mainly by changes in the sample volume, and is not an intrinsic luminosity dependence effect.

The density-independent estimates show much smaller differences in galaxy clustering with luminosity; the low-luminosity subsample (b) is slightly less strongly clustered than the other subsamples; there is no significant difference between the middle (c) and high (d) luminosity samples. Though these variations are smaller than those seen between the density-dependent estimates, the stability of the density-independent estimates means that they are much more significant. The slightly steeper and lower amplitude correlation function of faint galaxies compared to bright galaxies seems to be a real luminosity dependence effect.

In Figure 2 we also present $\xi(s)$ measured from samples (e) and (f), early and late type galaxies. As expected, the early types show significantly stronger clustering than late types. Due to a bias against classifying galaxies at large distances in the survey as early-type (reflected by the low value of $\langle V/V_{\max}\rangle$ in Table 1 of Paper 1), rather than generate the random $N(x)$ distribution from the measured luminosity function for galaxies of the appropriate type, we have instead fitted a fourth-order polynomial to the observed $N(x)$ (Fig. 1). For this reason, no estimates of the observed/expected density ratio $n_g/n_e$ are given in Table 1 for these two subsamples. The 'tail' in the density-dependent estimate of $\xi(s) \approx 0.1$ for early type galaxies is almost certainly due to the low $\langle V/V_{\max}\rangle$ for this sample.

Power-law fits, $\xi(s) = (s/s_0)^{-\gamma_s}$, to the density-independent estimates over the range 1.5–30 $h^{-1}\text{Mpc}$ are given in Table 1 for each sample. For sample (a), all galaxies, we find a power-law index $\gamma_s = 1.47$, shallower than $\gamma \approx 1.7$ measured in real space (e.g. Davis and Peebles 1983, Bean et al. 1983) and as determined from the angular correlation function $w(\theta)$ (e.g. Groth and Peebles 1977, Maddox et al. 1990c). This difference is due to redshift-space distortions (cf. the following section). Our estimates of $\gamma_s$ and $s_0$ are slightly less certain than earlier determinations since our sparse-sampling strategy was designed to minimise errors in $\xi$ on scales $s \approx 20h^{-1}\text{Mpc}$, where the amplitude of $\xi$ is low, not on small scales $s \approx s_0$.

The power-law index $\gamma_s$ becomes progressively steeper for lower luminosity galaxies, changing from 1.41 for the brightest sample to to 1.80 for the faintest sample. The correlation length for the lowest luminosity galaxies is $s_0 = 4.9h^{-1}\text{Mpc}$, slightly lower than for higher luminosities which have $s_0 \sim 6h^{-1}\text{Mpc}$. This corresponds to a factor of 1.7 in the correlation amplitude at $10h^{-1}\text{Mpc}$, but redshift-space distortions mean that this cannot be interpreted directly in terms of the relative bias factors.

Early type galaxies show a very shallow $\gamma_s \approx 1.25$ due to the large amplitude of $\xi$ on scales $\approx$ 10–30$h^{-1}$Mpc. As mentioned in Section 4, we believe this shallow slope is due to redshift distortions. The correlation length $s_0 = 9.6$ is significantly larger than for late type galaxies, corresponding to a ratio of 2.4 in amplitude at $10h^{-1}$Mpc; again redshift-space distortions must be taken into account in order to relate this value to relative bias factors.

## 4 The Real Space Correlation Function $\xi(r)$

The clustering results presented in the preceding section are of course affected by redshift-space distortions to uniform Hubble flow. On small scales, random peculiar velocities will cause clustering to be underestimated, while on large scales coherent bulk-flows will cause the clustering amplitude to be overestimated (eg. Kaiser 1987). Moreover, we expect that early type galaxies, which are preferentially found in high density regions, will be more strongly affected by redshift-space distortions than late type galaxies found in the field.

In order to measure clustering unaffected by redshift-space distortions, one must somehow integrate, or project, over the radial distance coordinate. There are several ways of doing this.

One can measure the angular correlation function $w(\theta)$ and invert it using Limber's equation. The problem with this method is that $w(\theta)$ measured from a redshift survey is noisy compared to the $w(\theta)$ that can be measured from a much larger photometric catalogue without the redundant redshift information. Additionally, the inversion is sensitive to the selection function for the relevant galaxy type. Due to the difficulty of classifying 17th mag galaxies on Schmidt plates, the luminosity functions for early and late-type galaxies in this survey are subject to large errors. As described in §4.1 we circumvent these problems by measuring $w(\theta)$ for the fully-sampled APM Bright Galaxy Catalogue (Loveday 1989) and estimating the selection function $S(z)$ by smoothing the observed $N(z)$ distribution for galaxies of appropriate type in the Stromlo-APM survey.

Another approach, followed by Davis and Peebles (1983), is to calculate the full redshift space correlation function $\xi(\sigma, \pi)$ as a function of the two components of separation parallel ($\pi$) and perpendicular ($\sigma$) to the line of sight, perform the integral $\Xi(\sigma) = \int_{-\infty}^{+\infty} \xi(\sigma, \pi) d\pi$ and then to invert the resulting projected correlation function $\Xi(\sigma)$ to obtain $\xi(r)$. However, since $\xi(\sigma, \pi)$ is now calculated on a 2d grid, it suffers from shot-noise due to the small number of galaxy-galaxy pairs in each $(\sigma, \pi)$ bin and so is very noisy. This is especially serious for our sparse-sampled redshift survey on small scales, and so this method is not used here.

A third method, when one has a sparse-sampled redshift survey drawn from a larger parent catalogue, is to calculate the projected cross-correlation between the redshift survey and its 2d parent survey. This projected correlation function is easily inverted to give $\xi(r)$. This method gives the most stable and reliable estimates of $\xi(r)$, and we apply it to our survey in §4.2.

### 4.1 Inversion of $w(\theta)$

In constructing the Stromlo-APM survey, Loveday (1989) inspected all APM galaxy candidates brighter than $b_J = 16.57$ and assigned each galaxy a morphological type. There are 4439 early type and 8844 late type galaxies in the APM Bright Galaxy Catalogue (APMBGC) and so we can calculate $w(\theta)$ for early and late type galaxies much more accurately from the APMBGC than from the 1:20 sparse-sampled redshift survey. As an additional bonus, the brighter mag limit of the APMBGC (16.57 vs 17.15) means that galaxy typing should be more reliable and complete (only 164 out of 13447 unmerged APMBGC galaxies are not classified as early or late type). Of course, redshifts, and hence luminosities, are not known for the vast majority of APMBGC galaxies, and so this analysis cannot be applied to the luminosity-selected samples.

We have estimated $w(\theta)$ using the estimator

$$w(\theta) = \frac{N_{gg}(\theta) N_{rr}(\theta)}{[N_{gr}(\theta)]^2} - 1 + \Delta w, \qquad (5)$$

where $N_{gg}$, $N_{gr}$ and $N_{rr}$ are the number of galaxy-galaxy, galaxy-random and random-random pairs at angular separation $\theta$ and $\Delta w$ is a correction for the integral constraint,

$$\Delta w = \int\int_{\text{survey}} w(\theta_{12}) d\Omega_1 d\Omega_2, \qquad (6)$$

(Groth and Peebles 1977). The correction $\Delta w$ is estimated in practice by calculating $w(\theta)$ without the correction, integrating $w(\theta)$ over all elements of solid angle $d\Omega_i$ in the survey area to obtain $\Delta w$ and recalculating $w(\theta)$ with the correction added. A stable solution is rapidly reached by iteration.

Fig. 4 shows $w(\theta)$ for all, early and late-type galaxies in APMBGC. We have fitted a power law $w(\theta) = A\theta^{1-\gamma}$ from 0.1 to 5° to these estimates, with results shown in Table 2. We see that early-type galaxies have a *steeper* power law slope than late-types, in agreement with Davis and Geller (1976) and Giovanelli *et al.* (1986) but contrasting with the redshift-space measurements (Table 1). The integral constraint corrections $\Delta w$ shown in Table 2 make a negligible difference to power-law fits on scales smaller than 5° but they do give some idea of possible systematic errors in the $w(\theta)$ estimates on large scales.

We have used these power law solutions in the relativistic version of Limber's equation (Groth and Peebles 1977, Phillipps *et al.* 1978) assuming $q_0 = 0.5$. The selection function $S(z)$ used in Limber's equation was determined separately for each galaxy type by smoothing the observed $N(z)$ for galaxies in the Stromlo-APM survey of the appropriate type and with $b_J < 16.57$ with a gaussian of FWHM = 0.01. The resulting parameters $r_0$ and $B$ for the spatial correlation function $\xi(r) = (r/r_0)^{-\gamma} = Br^{-\gamma}$ are shown in Table 2. We see that at $r = 1h^{-1}$Mpc, the clustering amplitude of early-type galaxies is more than a factor of three higher than that of late-type galaxies.

## 4.2 Projected Cross-Correlation $\Xi(\sigma)$

### 4.2.1 Method

Probably the most reliable way of determining the real-space correlation function when one has a sparsely-sampled redshift survey drawn from a fully sampled parent catalogue is to calculate the projected cross-correlation function $\Xi(\sigma)$ between the redshift survey and its 2d parent catalogue,

$$\Xi(\sigma) = \int_{-\infty}^{+\infty} \xi(\sqrt{\Delta y^2 + \sigma^2}) d\Delta y, \qquad (7)$$

where the integral extends over all line-of-sight separations $\Delta y$ for pairs of galaxies with constant projected separation $\sigma = y\theta$ ($\theta$ is the angular separation and $y$ is the distance to the galaxy in the redshift survey). This projected function can be directly inverted numerically to give a stable estimate of $\xi(r)$, which is unaffected by redshift-space distortions. This method was used by Saunders et al. (1992) to measure $\xi(r)$ for IRAS galaxies using the QDOT redshift survey (Lawrence et al., 1994) and was earlier used by Lilje and Efstathiou (1987) to measure the cross-correlation of Lick galaxies with Abell clusters.

Here, we cross-correlate the redshift survey samples listed in Table 1 with 36,276 galaxies brighter than $b_J = 17.15$ in the APM Galaxy Survey. The resulting estimate of $\xi$ uses all galaxies from the parent sample as well as the sparse subsample with measured redshifts and so minimises random errors and enables us to estimate $\xi(r)$ on much smaller scales than possible using the redshift survey galaxies alone. A further advantage of this procedure for our analyses is that the estimates of $\xi$ are independent of uncertainties in the selection function for galaxies of a specific type.

To estimate $\Xi$ we consider each redshift survey galaxy at known distance $y$, and count the number of APM galaxies $N_g(\sigma)$ at projected separation $\sigma = y\theta$ and then compare this with the number of random points $N_r(\sigma)$ (scaled by the relative numbers of galaxies and random points) at the same separation.

An estimate of the projected correlation function,

$$X_i(\sigma) = \frac{1}{p(\sigma, y)} \left[ \frac{N_g(\sigma)}{N_r(\sigma)} - 1 \right] + \Delta X_i, \qquad (8)$$

is thus calculated for each redshift survey galaxy, where the factor

$$p(\sigma, y) = \frac{1}{\bar{N}\Xi(\sigma)} \int_0^\infty \psi(x) x^2 \xi(r) dx, \quad r^2 = x^2 + y^2 - 2xy\cos(\sigma/y) \qquad (9)$$

corrects for projection effects and biases introduced by assuming that $\xi(r)$ is negligible on scales $r \sim y$, $\psi(x)$ is the galaxy density at distance $x$ and $\bar{N}$ is the surface density of galaxies in the 2d catalogue. The term $\Delta X_i$ is a correction for the integral constraint affecting each redshift survey galaxy. It is estimated by assuming a truncated power-law model $\Xi(\sigma) = (\sigma/\sigma_0)^{-\alpha}$ for $\sigma < \sigma_{max}$, zero otherwise, and integrating over all solid angle elements in the survey area,

$$\Delta X_i = \int \int_{\text{survey}} \Xi(\sigma_{12}) d\Omega_1 d\Omega_2. \qquad (10)$$

We assume parameters $\alpha = 0.81$, $\sigma_0 = 165 h^{-1}$Mpc and $\sigma_{\max} = 20 h^{-1}$Mpc, which give a reasonable fit to the final $\Xi(\sigma)$ obtained from the all-galaxy sample. The correction $\Delta X_i$ varies with the distance $y$ of the redshift survey galaxy, from $\approx 0.1$ at $y \approx 300 h^{-1}$Mpc to $\approx 20$ at $y = 10 h^{-1}$Mpc.

Saunders et al. (1992) made an estimate $X_i(\sigma)$ of $\Xi(\sigma)$ for each redshift survey galaxy at distance $y_i$, and then formed a weighted average of the $X_i$ to obtain a final $\Xi(\sigma)$. In order to mimimise the shot-noise in each $X_i(\sigma)$ estimate, we instead chose to bin the redshift survey galaxies into distance bins of width $\Delta y = 10 h^{-1}$Mpc and then estimated $X_i$ for each bin centred on distance $y_i$.

Estimating the $X_i$ in an unbiased way requires the correct values for $p(\sigma, y)$ and $\Delta X_i$ (Eqn. 8), which in turn require the prior knowledge of $\xi(r)$ and $\Xi(\sigma)$ (Eqns. 9 and 10). Fortunately $p$ and $\Delta X_i$ are only weakly dependent on $\xi$ and $\Xi$ and so an unbiased solution can be calculated by iteration from approximate initial estimates of $\xi$ and $\Xi$. We assume initial power-law forms for $\xi(r)$ and $\Xi(\sigma)$ and calculate $X_i$ for each distance bin $y_i$. Then we take a weighted average of the $X_i$'s using weights designed to give the minimum-variance estimate of $\Xi$ (Saunders et al eq. 16). This estimate of $\Xi(\sigma)$ is then numerically inverted using Eq. 26 of Saunders et al to obtain an estimate of $\xi(r)$. Note that this inversion does *not* assume a power-law form for $\xi(r)$, an assumption which can lead to a systematic bias in the slope $\gamma$ (Saunders 1994). Power laws are separately fitted to the projected and spatial correlation functions over scales 0.2–10 $h^{-1}$Mpc and the process is iterated to obtain a stable solution. Typically five iterations are required.

### 4.2.2 Test of the Method

In order to test the method, we have used one of the CDM-like N-body simulations of Croft and Efstathiou (1994). These simulations combine a large volume (box length = $300h^{-1}$Mpc) with a spatial resolution of $\approx 80h^{-1}$kpc, and so can be used to generate reasonable approximations to our redshift survey. We analysed a simulation from ensemble B, a model with non-zero cosmological constant $\lambda = \Lambda/(3H_0^2) = (1 - \Omega_0) = 0.8$. Weights were assigned to each particle using the peak background split algorithm (Bardeen *et al.* 1986, White *et al.* 1987) and 'galaxies' selected within the APM area and with the Stromlo-APM selection function (Paper 1).

In Fig. 5 (solid symbols), we plot the correlation function $\xi(r)$ measured directly from the real-space positions of the 34,120 'galaxies' in the simulation. Over separations 0.2–20 $h^{-1}$Mpc, $\xi(r)$ is very well fit by a power-law $\xi(r) = (r/r_0)^{-\gamma}$ with $\gamma = 1.89$ and $r_0 = 6.2h^{-1}$Mpc.

We then generated five sub-samples from this simulation. For each sub-sample, galaxies were selected from the full simulation at random with probability 0.05, the same sparse-sampling rate used for the Stromlo-APM Survey. Each of these five mock redshift surveys was cross-correlated, using the *redshift*-distance information, with the sky-projected data from the full simulation. Estimates of $\Xi(\sigma)$ and thence $\xi(r)$ were determined for each mock redshift survey using the method described above. In Fig. 5, the average of the $\xi(r)$ estimates from the cross-correlation procedure is shown by the open symbols and the error bars show the scatter between estimates. It can be seen that this estimate of $\xi(r)$ gives excellent agreement with the 'true' $\xi(r)$ from the full simulation. The increased noise on scales $r \gtrsim 20h^{-1}$Mpc is not surprising since the power-law fits to $\Xi(\sigma)$ and $\xi(r)$ used in (9) were only made over the range 0.2–10 $h^{-1}$Mpc. A final power-law fit to the average $\xi(r)$ gives $\gamma = 1.94 \pm 0.08$ and $r_0 = 6.1 \pm 0.4$, consistent with the direct determination of $\xi(r)$ from the full simulation.

### 4.2.3 Comparisons Between Galaxy Samples

Our estimates of $\xi(r)$ obtained by inversion of the projected cross-correlation function between all APM galaxies in the parent sample and each of the subsamples of the redshift survey listed in Table 1 are shown in Fig. 6. Power-law fits over the range $0.2 < r < 20h^{-1}$Mpc are shown by dashed lines and the values of the slope $\gamma_r$ and correlation length $r_0$ are listed in Table 1, where again the quoted errors are obtained from the variance between nine bootstrap resamplings of the redshift survey sample. To aid comparing the samples, the dotted line in each panel of Fig. 6 shows $\xi(r)$ predicted in a $\Gamma = 0.2$ biased CDM model.

For sample (a), all galaxies, we determine a correlation function slope $\gamma_r = 1.71 \pm 0.05$ and a correlation length $r_0 = 5.1 \pm 0.2$, in good agreement with earlier determinations (*e.g.* Davis and Peebles 1983).

We see no significant difference between the clustering of $L^*$ galaxies (sample c) and brighter galaxies (sample d). This is perhaps a surprising result given some previous work (*e.g.* Hamilton 1988), but is consistent with our density-independent estimates of $\xi$ in redshift space (Section 3). We find that intrinsically faint galaxies (sample b) have a much steeper correlation function slope, $\gamma_r = 2.09$, and smaller correlation length, $r_0 = 3.2$, compared to $L^*$ and brighter galaxies (samples c & d). This comfirms the trends which are seen at lower significance in our redshift-space estimates of Section 3. The correlation amplitude for sample (b) is a factor $\sim 2$ lower than (c) and (d) at $1h^{-1}$Mpc and the factor increases to $\sim 4$ at $10h^{-1}$Mpc. A steeper slope might be expected if low-luminosity galaxies are found mostly in cluster environments and thus co-habit with early-type galaxies, as suggested by the steeper luminosity function faint-end slope found in cluster environments compared with the field (eg. Binggeli *et al.* 1988).

We see that for early-type galaxies (sample e) the slope is $\gamma_r = 1.85$, steeper than for late-type galaxies (sample f) which have $\gamma_r = 1.64$. These slopes and also the corresponding scale lengths are in good agreement with observations of $w(\theta)$ (Table 2). This confirms that the very shallow slope ($\gamma_s = 1.25$) seen in the redshift space estimate of $\xi$ for early type galaxies is due to redshift-space distortions.

It is important to remember that the $\xi(r)$ estimates shown in Fig. 6 are for the *cross*-correlation of galaxies of specified type with galaxies of all types. The differences between the cross-correlation functions of the samples should be smaller than the differences between the *auto*-correlation functions. For example, suppose that the mean galaxy bias relative to mass is $b$, and that early and late type galaxies have bias $b_e$ and $b_l$. The cross-correlation functions $\xi_{eg}$ and $\xi_{lg}$ will have amplitudes $B_{eg} \propto b_e \times b$ and $B_{lg} \propto b_l \times b$, with ratio $b_e/b_l$. The auto-correlation functions $\xi_{ee}$ and $\xi_{ll}$ will have amplitudes $B_{ee} \propto b_e^2$ and $B_{ll} \propto b_l^2$, with ratio $(b_e/b_l)^2$. From Table 1, the ratio $\xi_{eg}/\xi_{lg} \approx 2.3$ at $1h^{-1}$Mpc. From Table 2, we see that the ratio $\xi_{ee}/\xi_{ll} \approx 3.5$ and not 5.5 as expected. Formally this discrepancy is marginally significant compared to the estimated errors, but note that the magnitude limit used to estimate $w$ is different to that used for $\Xi$ and so different volumes are being sampled. Also the amplitude $B$ is sensitive to the power-law slope $\gamma$ used in the Limber inversion. As discussed by Saunders (1994), the fact that $w(\theta)$ is not a pure power-law, but contains a break, can result in a systematic bias in the 'power-law' slope $\gamma$.

### 4.2.4 Effect of Sampling Fluctuations

One might argue that the observed segregation at low luminosities and lack of segregation at high luminosities could be due to sampling fluctuations, since, however ingenious our estimator, low-luminosity galaxies are necessarily closer to us than more luminous galaxies. Ideally, one would like to form a volume-limited sample containing galaxies over a range of luminosities at uniform radial density, and so calculate the clustering of galaxies of different luminosities within the same volume. Unfortunately, true volume-limited samples drawn from the Stromlo-APM survey are very small indeed. This is because of the difficulty of photographic calibration brighter than 15th magnitude — a strictly volume limited sample must have a minimum distance and upper luminosity limit, as well as a maximum distance and lower luminosity limit.

By way of compromise, we have calculated the spatial cross-correlation functions for the luminosity-selected samples with additional distance limits imposed. In Fig. 7a, we plot $\xi(r)$ for the faint (solid symbols) and middle (open symbols) luminosity galaxies, limiting to galaxies in the range 70–140 $h^{-1}$Mpc (the range of overlap for these two samples — see Fig. 1). We see that even when drawn from the same volume, middle-luminosity galaxies still show stronger clustering than low-luminosity galaxies. In Fig. 7b, we compare the clustering of middle (solid symbols) and high (open symbols) luminosity galaxies in their range of overlap (90–220 $h^{-1}$Mpc). No clear difference in clustering is seen, in agreement with Fig. 6, in which no explicit volume constraint was imposed. Finally, in Fig. 7c, we plot the clustering of middle-luminosity galaxies in the two different volumes: 70–140 $h^{-1}$Mpc (solid symbols) and 90–220 $h^{-1}$Mpc (open symbols). It is indeed encouraging that the same luminosity galaxies have the same measured clustering in two different (albeit not entirely independent) volumes. We thus believe that the observed segregation at low-luminosities and lack of segregation at high-luminosities is a genuine effect, and is not due to sampling fluctuations in our survey.

### 4.2.5 Separating Luminosity and Morphological Segregation

One might also ask whether the differences we see between the clustering of faint and middle luminosity samples are in fact due to true luminosity segregation or just a different balance of morphological types at different luminosity. Conversely, if, as Hasegawa and Umemura (1993) claim, early and late type galaxies show luminosity segregation of opposite sign, then the lack of segregation between the middle and bright luminosity samples could be due to cancelling of effects for early and late types.

Since our $\xi(r)$ cross-correlation estimates in Fig. 6 have such small error bars, it is worth investigating the dependence of clustering on morphology and luminosity separately, ie. by further dividing the early and late type galaxies by luminosity. The cross-correlation $\xi(r)$ measured for these new samples are shown in Fig. 8 and the results of power-law fits from 0.2 to $20h^{-1}$Mpc are given in Table 3.

Both early and late type galaxies separately show evidence of luminosity segregation between the faint and middle luminosity samples. The signal for the faintest ($-19 < M < -15$) early-type galaxies goes negative on scales $\approx 6$–$20h^{-1}$Mpc and so for this subsample, the power-law fit was truncated at $5h^{-1}$Mpc. Evidence for luminosity segregation brighter than $L^*$ is marginal at best, although possibly the brightest late type galaxies show slightly enhanced large-scale clustering over $L^*$ galaxies. The steeper, higher amplitude clustering of early-type compared to late-type galaxies occurs for all luminosity classes except for the faintest one, which gives a rather noisy correlation function for early types.

The results presented in fig. 8 thus show that luminosity and morphological segregation are both real, independent effects.

## 5 Discussion and Conclusions

We have presented estimates for the correlation function of various galaxy samples in redshift space using two different estimators. The differences in redshift space correlation functions determined using density-dependent and density-independent estimators highlight the problems in trying to determine clustering from a sample in which the mean density is not well-defined. We have shown that the density-independent estimator (3) provides a more reliable determination of galaxy clustering when analysing subsamples of a catalogue in which the actual galaxy density differs from the expected density, or when the selection function determined from the observed luminosity function does not provide a perfect fit to the observed radial density.

We have seen that the redshift-space correlation function is significantly affected by peculiar velocities and present estimates of $\xi(r)$ unaffected by these distortions. We find that early-type galaxies show a steeper correlation function slope and larger correlation length than late type galaxies. Low-luminosity galaxies exhibit a steeper slope and smaller correlation length than $L^*$ galaxies, but no significant difference is seen in the clustering of $L^*$ and super-$L^*$ galaxies.

Our results concerning morphological segregation of galaxies are consistent with earlier investigations by for example Davis and Geller (1976), Giovanelli *et al.* (1986) and Iovino *et al.* (1993), who all find that early type galaxies are significantly more strongly clustered, and with a steeper correlation function slope, than late type galaxies.

Given our results for variation of clustering strength with luminosity, it is not too surprising that previous analyses using smaller samples have not all agreed on the existence of luminosity segregation. Our results are consistent with the majority of analyses which found no luminosity segregation or only a small difference in the clustering of faint and bright galaxies (see references in §1).

Our results are not consistent with those of Hamilton (1988) who finds significantly enhanced clustering of the brightest galaxies compared with $L^*$ galaxies. Hamilton devised a test for luminosity segregation insensitive to variations in galaxy density by comparing clustering of galaxies of different luminosity in the same volume. The correlation function as a function of absolute magnitude is built up by multiplying ratios of correlation functions measured in successive volume-limited samples of the data. Unfortunately, this technique also accumulates errors in $\xi$ as one works away from the fiducial luminosity, and so it is hard to assess the significance of the apparently enhanced clustering of the most luminous galaxies seen by Hamilton. Interestingly, when Hasegawa and Umemura (1993) repeated Hamilton's analysis after correcting the CfA magnitudes for internal and galactic obscuration, the luminosity effect is much weakened.

In most previous studies, low luminosity galaxy samples have been dominated by Virgo and the local supercluster, and so even if one allows for variation in galaxy density between samples (eg. Davis *et al.* 1988, Maurogordato and Lachieze-Rey 1991), one is still comparing clustering of faint galaxies in one small volume of the Universe with bright galaxies drawn from a much larger volume. Moreover, one expects galaxy peculiar velocities to be larger in high density regions such as the local supercluster, and hence redshift-space distortions may have had a stronger effect on low-luminosity galaxy samples than on high-luminosity ones.

The significantly fainter apparent magnitude limit of the Stromlo-APM survey compared with earlier surveys means that our low-luminosity sample is drawn from a much larger volume of the Universe than was possible before — even our lowest luminosity sample has a median depth of $\sim 100 h^{-1}$Mpc. Therefore the statistical fluctuations on our clustering measurements for faint galaxies should be small, enabling reliable comparison with high-luminosity galaxy samples. Indeed, we have demonstrated this by comparing clustering of different luminosity galaxies in the same volume.

The observed variation of galaxy clustering with morphological type and the observed weaker clustering of low-luminosity galaxies is what one would expect in biased galaxy formation scenarios. However, the lack of luminosity segregation at brighter luminosities is not compatible with some simple theories of biased galaxy formation, such as "natural bias" (White *et al.* 1987, Valls-Gabaud *et al.* 1989) in which galaxies preferentially form in the peaks of the underlying mass fluctuations.

The mean absolute magnitudes for our three luminosity subsamples are $-18.4$, $-19.6$ and $-20.6$. According to the White *et al.* model, and using the relation (1) between circular velocity and absolute blue magnitude given by White, Tully and Davis (1988), one would expect enhancements in the amplitude of $\xi(r)$ by roughly a factor of 1.5 from each luminosity subsample to the next. Instead, we see an enhancement by factor 2–4 between the first two samples, and no significant difference between the second two samples. Conceivably, a closely related biasing model, but modified by a feedback mechanism at high luminosities, might be able to explain our observations.

In order to compare the correlation functions for the various samples in real and redshift space more directly, in Fig. 9 we plot the real-space ($\xi(r)$, solid symbols) and redshift-space ($\xi(s)$, open symbols) correlation functions on the same plot. We have also re-binned the galaxy pair counts into coarser separation bins in order to reduce the error bars. For the all galaxies sample (a), we see a surprisingly small difference between $\xi(r)$ and $\xi(s)$. This suggests a relatively low value for the quantity $\beta = \Omega^{0.6}/b \sim 0.3$, where $\Omega$ is the cosmological density parameter and $b$ is the bias parameter for optically selected galaxies. Further discussion of this topic is postponed until the next paper in this series (Loveday *et al.* 1995).

Note that the ratio $\xi(s)/\xi(r)$ in the linear regime for different samples will differ due to both 1) varying amplitude of redshift-space distortion with changes in the bias parameter $b_t$ for the different samples and 2) differences between the cross and auto-correlation functions. As discussed in §4.2.3, an auto-correlation function will scale as $b_t^2$ whereas a cross-correlation function will only scale as $b_t$. These two effects pull the $\xi(s)$ auto-correlation function in opposite directions relative to the $\xi(r)$ cross-correlation function and so the large differences seen for low-luminosity galaxies (b) and early-type galaxies (e) are quite surprising. A quantitative analysis of this problem is again deferred to Loveday *et al.* (1995).

It is intriguing that morphological segregation is strongest on scales $\lesssim 1h^{-1}$Mpc whereas luminosity segregation is strongest on scales $\gtrsim 1h^{-1}$Mpc, hinting that the weaker clustering of low-luminosity objects may be a purely primordial (biasing) effect, but that morphological segregation may be enhanced by environmental effects, such as galaxy interactions and merging. Well motivated and well specified models of biasing as well as more observational data are needed to make further progress in understanding morphological and luminosity segregation, and thus providing an important key to unlocking the secrets of galaxy formation.

**Acknowledgements** We thank Andrew Hamilton and Will Saunders for useful discussions.

# Appendix: Optimal Weighting of Galaxies in a Redshift Survey for Estimating $\xi$ on Large Scales

Imagine a homogeneous catalogue of galaxies with a well defined mean density $n$. Assuming that $\xi(r) \ll 1$, then a realistic estimate of the error in $\xi$ is given by

$$\delta \xi(r) \approx \frac{1 + 4\pi n J_3}{\sqrt{N_p}}, \tag{11}$$

where,

$$J_3(r) = \int_0^r x^2 \xi(x) dx, \tag{12}$$

and $N_p$ is the number of galaxy pairs used to estimate $\xi(r)$ for this separation bin (Peebles 1973, Kaiser 1986). Essentially, this model assumes that galaxies occurr in clusters of $N_c = 1 + 4\pi n J_3$ members, and so the number of *independent* pairs in an estimate of $\xi(r)$ is given by $N_p/N_c^2$.

Eqn. 11 will give an accurate estimate of $\delta \xi$ for a homogeneous sample—*e.g.* an N-body simulation, but real redshift surveys are *not* homogeneous—the observed number density $n$ of galaxies decreases with distance $x$ from the observer. How should Eqn. 11 be applied in this case, and how should we weight each galaxy to minimise $\delta \xi$ on large scales?

We can estimate an optimum weighting scheme by considering galaxies in concentric shells about the observer, whose width is much larger than the correlation length $r_0$. Suppose there are $M$ '$x$-shells' of width $l \gg r_0$ and volume $V_i$ centred on the observer. If the density of galaxies in shell $i$ is $n_i$, we expect $V_i n_i$ galaxies in this $x$-shell. Around each galaxy, neglecting edge effects, we estimate $\xi(r)$ in a spherical shell ('$r$-shell') of volume $\Delta V = 4\pi r^2 \Delta r$. If $r \ll x$, then in the absence of clustering we would expect each $r$-shell to contain $\Delta V n_i$ galaxies. Therefore the expected number of pairs at separation $r \pm \Delta r/2$ in $x$-shell $i$ is $\langle N_{p_i} \rangle = n_i^2 V_i \Delta V$. If we actually count $N_{p_i}$ pairs, then our estimate of $\xi(r)$, which we denote $\xi_e(r)$, is given by

$$1 + \xi_e(r) = \frac{\sum_{i=1}^{M} W_i^2 N_{p_i}}{\sum_{i=1}^{M} W_i^2 n_i^2 V_i \Delta V}, \tag{13}$$

where $W_i$ is a weight (to be determined) given to each galaxy in $x$-shell $i$, the sums run over the $M$ $x$-shells, and we neglect pairs that cross shells.

Now the error in $\xi$ for shell $i$ is

$$\delta \xi = \frac{\delta N_{p_i}}{\langle N_{p_i} \rangle}, \tag{14}$$

and so from (11),

$$\delta N_{p_i} = (N_{p_i})^{1/2} (1 + 4\pi n_i J_3). \tag{15}$$

The variance in $\xi_e(r)$ is given by

$$\langle \delta \xi_e^2 \rangle = \frac{\sum_{i=1}^{M} W_i^4 N_{p_i} (1 + 4\pi n_i J_3)^2}{\left[ \sum_{i=1}^{M} W_i^2 n_i^2 V_i \Delta V \right]^2}. \tag{16}$$

Substituting $N_{p_i} = n_i^2 V_i \Delta V$ (assuming $\xi(r) \ll 1$), and replacing the sums over $x$-shells with integrals, we get

$$\langle \delta \xi_e^2 \rangle = \frac{1}{\Delta V} \frac{\int W^4 n(x)^2 [1 + 4\pi n(x) J_3]^2 dV}{\left[ \int W^2 n(x)^2 dV \right]^2}, \quad (17)$$

where $dV = \omega x^2 dx$ for a redshift survey cone of solid angle $\omega$, and the integration limits are the chosen distance limits for the redshift survey. Differentiating with respect to W, $\frac{\partial \langle \delta \xi^2 \rangle}{\partial W} = 0$ if

$$\int W^3 n(x)^2 [1 + 4\pi n(x) J_3]^2 dV \int W^2 n(x)^2 dV -$$
$$\int W n(x)^2 dV \int W^4 n(x)^2 [1 + 4\pi n(x) J_3]^2 dV = 0 \quad (18)$$

and so

$$W = \frac{1}{[1 + 4\pi n(x) J_3]}. \quad (19)$$

Of course, this optimal weighting requires prior knowledge of $J_3(r)$, but in practice $W$ is only weakly dependent on $J_3(r)$ if it is large enough, and a stable solution may quickly be reached by iteration.

Using this optimal weighting scheme, the estimated variance in $\xi$ for small $\xi$ is given by

$$\langle \delta \xi_e^2 \rangle = \frac{1}{\Delta V} \left[ \int_{x_{min}}^{x_{max}} \frac{n(x)^2 dV}{[1 + 4\pi n(x) J_3]^2} \right]^{-1}. \quad (20)$$

# Tables

Table 1: Sample definitions and properties

|   | Type  | $M_{\min}$ | $M_{\max}$ | $N_{\text{gal}}$ | $n_g/n_e$ | $\gamma_s$ | $s_0$ | $\gamma_r$ | $r_0$ |
|---|-------|------------|------------|------------------|-----------|------------|-------|------------|-------|
| a | All    | $-22$ | $-15$ | 1757 | 0.991 | $1.47 \pm 0.12$ | $5.9 \pm 0.3$ | $1.71 \pm 0.05$ | $5.1 \pm 0.2$ |
| b | Faint  | $-19$ | $-15$ |  473 | 0.965 | $1.80 \pm 0.18$ | $4.9 \pm 0.6$ | $2.09 \pm 0.13$ | $3.2 \pm 0.6$ |
| c | Middle | $-20$ | $-19$ |  661 | 1.154 | $1.60 \pm 0.22$ | $6.4 \pm 0.8$ | $1.66 \pm 0.06$ | $6.0 \pm 0.4$ |
| d | Bright | $-22$ | $-20$ |  544 | 0.952 | $1.41 \pm 0.31$ | $5.8 \pm 1.6$ | $1.79 \pm 0.11$ | $5.8 \pm 0.3$ |
| e | E&S0   | $-22$ | $-15$ |  336 | $\cdots$ | $1.25 \pm 0.33$ | $9.6 \pm 1.2$ | $1.85 \pm 0.13$ | $5.9 \pm 0.7$ |
| f | Sp&Irr | $-22$ | $-15$ | 1062 | $\cdots$ | $1.49 \pm 0.21$ | $5.3 \pm 0.4$ | $1.64 \pm 0.05$ | $4.4 \pm 0.1$ |

Notes.—$\gamma_s$ and $s_0$ are the power-law fit parameters to the correlation function in redshift space measured with the density-independent estimator over the range 1.5–30 $h^{-1}$Mpc. $\gamma_r$ and $r_0$ are the real-space power-law parameters over 0.2–20 $h^{-1}$Mpc determined from cross-correlation with the 2d APM survey (§4.2).

Table 2: Angular correlation function results for all, early and late type galaxies in APMBGC

| Type  | $\gamma$ | $A$ | $\Delta w$ | $B$ | $r_0$ |
|-------|----------|-----|------------|-----|-------|
| *All*   | $1.77 \pm 0.03$ | $0.22 \pm 0.01$ | $4.2 \times 10^{-3}$ | $19.0 \pm 0.8$ | $5.31 \pm 0.15$ |
| *Early* | $1.87 \pm 0.07$ | $0.40 \pm 0.03$ | $1.2 \times 10^{-2}$ | $46.0 \pm 5.0$ | $7.76 \pm 0.35$ |
| *Late*  | $1.72 \pm 0.05$ | $0.18 \pm 0.01$ | $3.4 \times 10^{-3}$ | $13.2 \pm 0.8$ | $4.49 \pm 0.13$ |

Note.—Power-law fits ($w = A\theta^{1-\gamma}$) were made over the range 0.1–5°. The integral constraint $\Delta w$ is estimated from the observed $w(\theta)$. The amplitude, $B$, and corresponding scale length, $r_0$ are for the spatial correlation function inferred from inverting Limber's equation.

Table 3: Results of power-law fits over 0.2–20 $h^{-1}$Mpc for cross-correlation of joint morphology-luminosity selected samples with the 2d APM survey.

| Type | | | Early Type | | | Late Type | | |
|------|------|------|------|------|------|------|------|------|
| $M_{\min}$ | $M_{\max}$ | $N_{\text{gal}}$ | $\gamma_r$ | $r_0$ | $N_{\text{gal}}$ | $\gamma_r$ | $r_0$ |
| $-19$ | $-15$ |  78 | $1.94 \pm 0.22$ | $3.4 \pm 0.8$ | 344 | $2.01 \pm 0.10$ | $2.9 \pm 0.4$ |
| $-20$ | $-19$ | 102 | $1.85 \pm 0.06$ | $6.4 \pm 0.7$ | 403 | $1.68 \pm 0.09$ | $4.9 \pm 0.2$ |
| $-22$ | $-20$ | 153 | $1.80 \pm 0.13$ | $7.4 \pm 1.1$ | 312 | $1.53 \pm 0.10$ | $6.2 \pm 0.4$ |

# Figure Captions

**Figure 1** $N(x)$ distributions for various galaxy samples as labelled. The dotted histograms show the predicted distributions from luminosity function estimates (a–d) and a fourth-order polynomial fit (e&f).

**Figure 2** The spatial correlation function $\xi(s)$ estimated from the Stromlo-APM redshift survey using density-dependent (open symbols) and density-independent (solid symbols) estimators for various samples as labelled. Error bars show the rms variance between nine bootstrap resamplings of the data. The dotted line shows the prediction of a $\Gamma = 0.2$ biased CDM model (LEPM), and is shown to aid in comparing the samples.

**Figure 3** The redshift-space correlation function $\xi(s)$ determined from a subsample of the survey data volume-limited to $200 h^{-1}$Mpc. Open symbols show results from the density-dependent estimator, solid symbols results from the density-independent estimator.

**Figure 4** The angular correlation functions $w(\theta)$ for all (open circles), early-type (filled circles) and late-type (filled squares) galaxies in the APM Bright Galaxy Catalogue. The dotted lines show power-law fits from 0.1 to $5°$.

**Figure 5** Test of the cross-correlation procedure. Solid symbols show $\xi(r)$ determined directly from a fully-sampled CDM-like N-body simulation; error bars are from the scatter between four zones. The open symbols show the average $\xi(r)$ determined from cross-correlating five 1:20 random samplings of the simulation with the 2d information from the full simulation; error bars show scatter between the five estimates.

**Figure 6** The real-space cross-correlation function $\xi(r)$ determined by inversion from the projected cross-correlation function $\Xi(\sigma)$ for the galaxy samples listed in Table 1. Error bars show the scatter between nine bootstrap resamplings of the Stromlo-APM catalogue.

**Figure 7** Distance limited cross-correlation functions. (a) 70–140 $h^{-1}$Mpc volume — low (solid symbols) and middle (open symbols) luminosity galaxies; (b) 90–220 $h^{-1}$Mpc volume — middle (solid symbols) and high (open symbols) luminosity galaxies; (c) middle-luminosity galaxies in the two different volumes: 70–140 $h^{-1}$Mpc (solid symbols) and 90-220 $h^{-1}$Mpc (open symbols).

**Figure 8** The real-space cross-correlation function $\xi(r)$ determined by inversion from the projected cross-correlation function $\Xi(\sigma)$ for galaxy samples selected by both morphology and luminosity.

**Figure 9** Comparison of real-space clustering (solid symbols) and redshift-space clustering (open symbols).

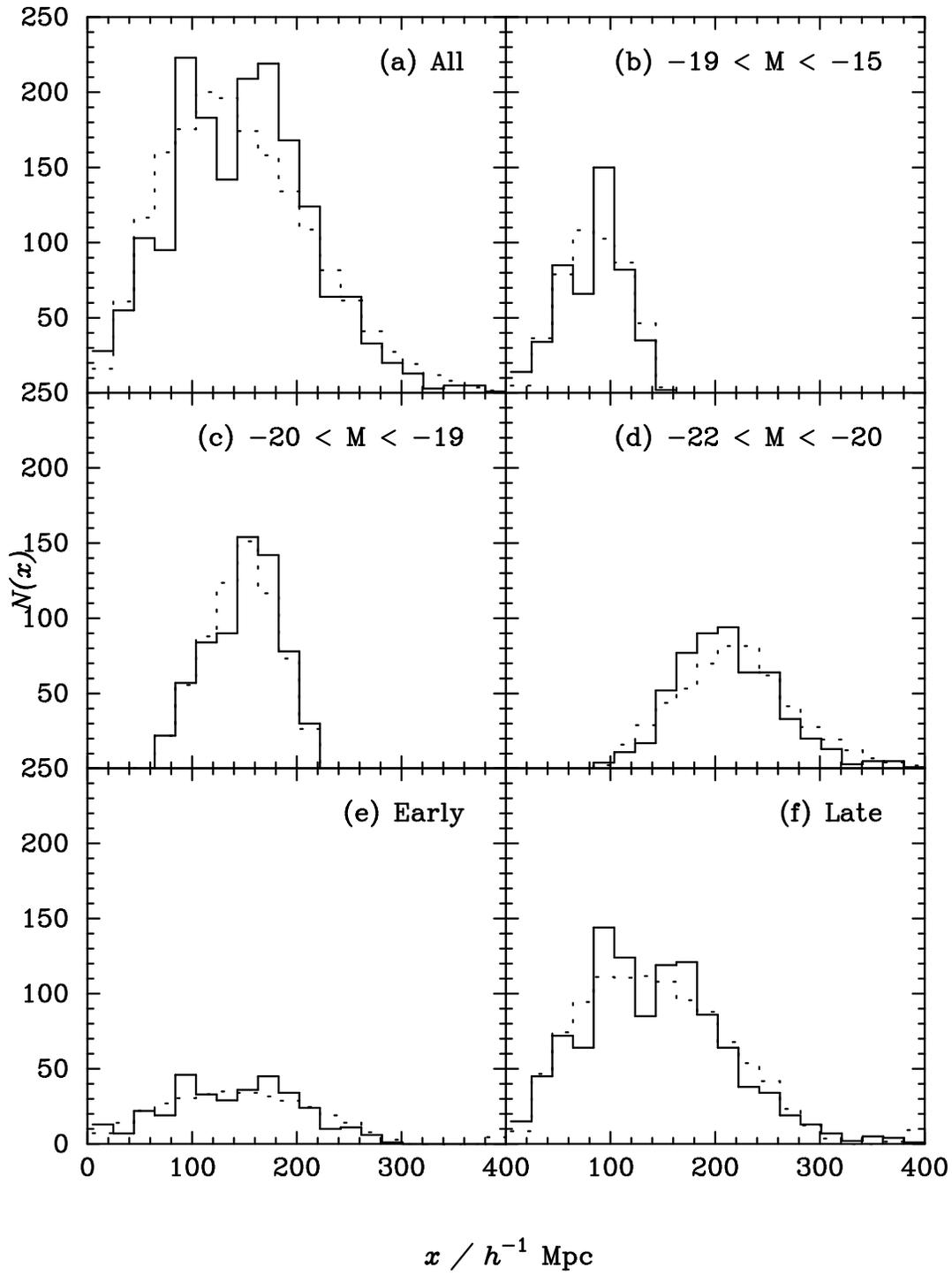

Figure 1:

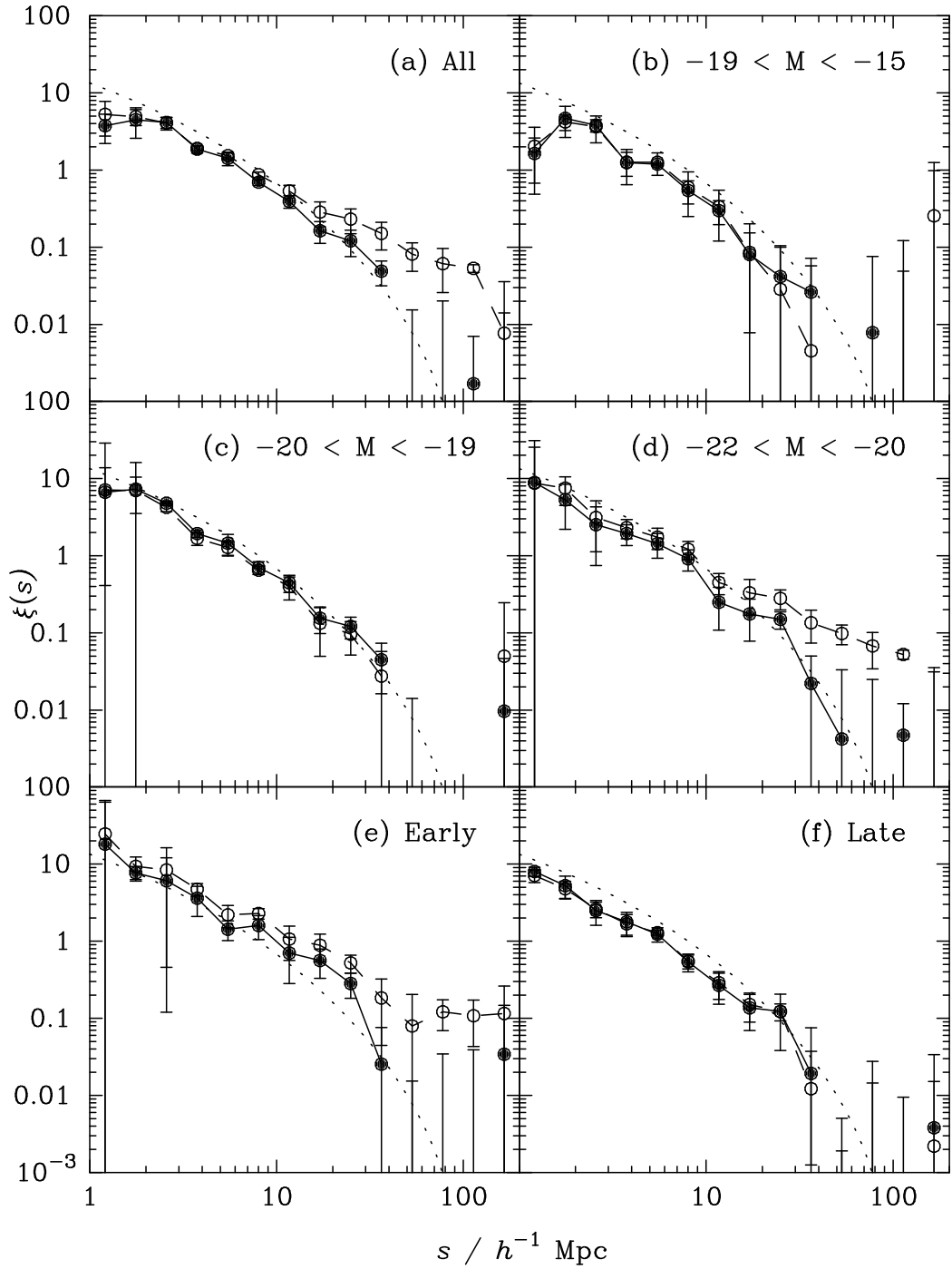

Figure 2:

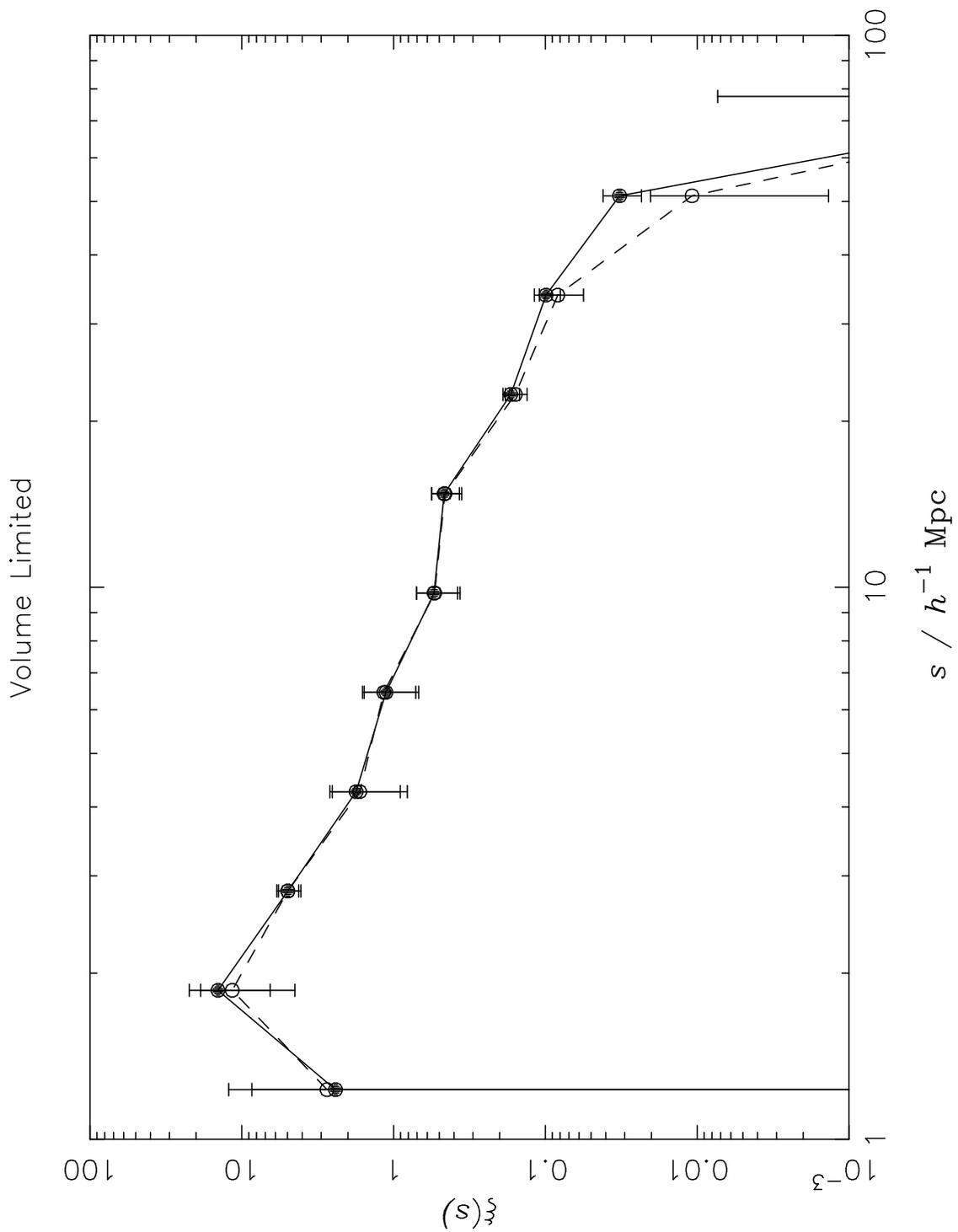

Figure 3:

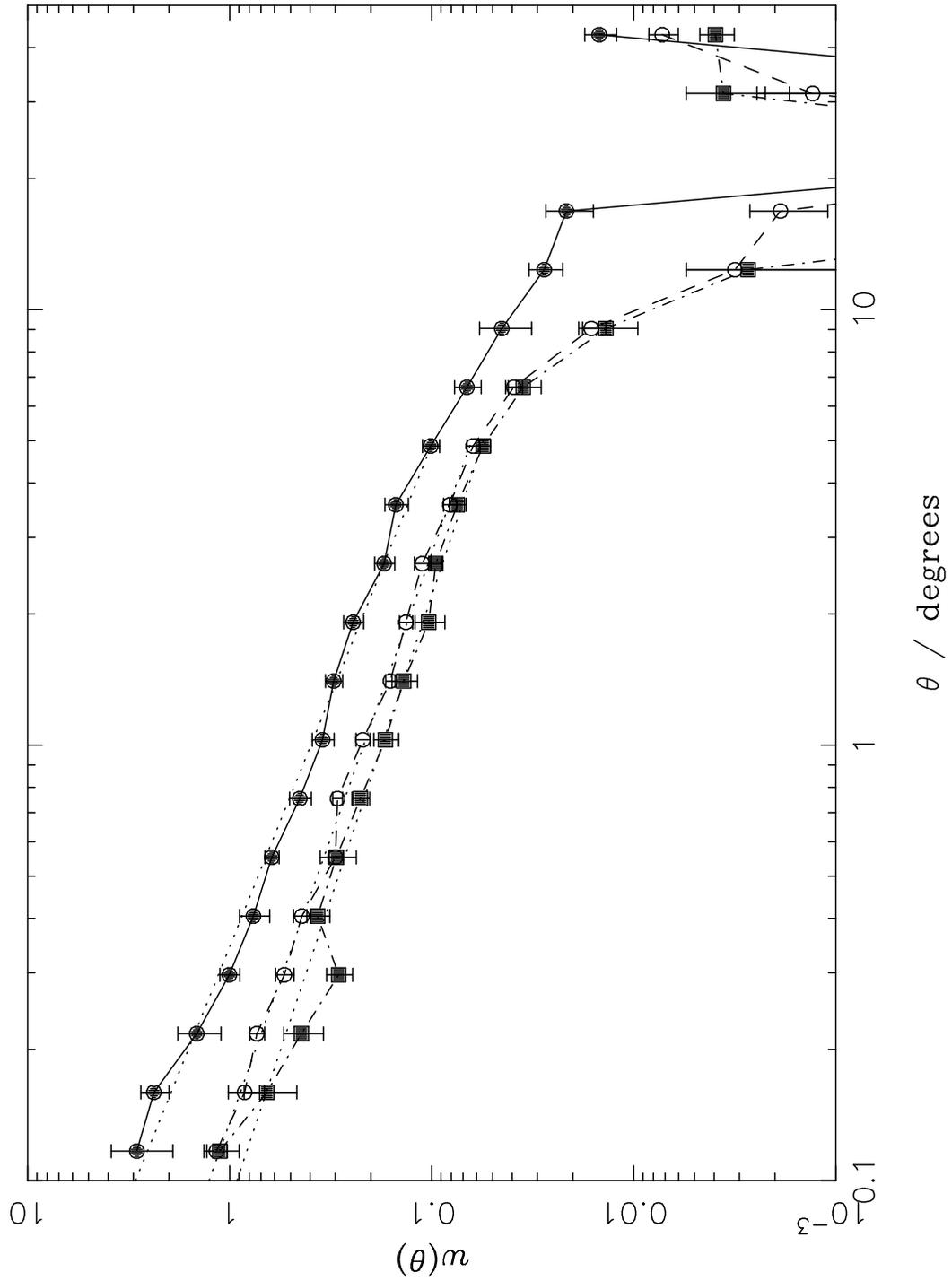

Figure 4:

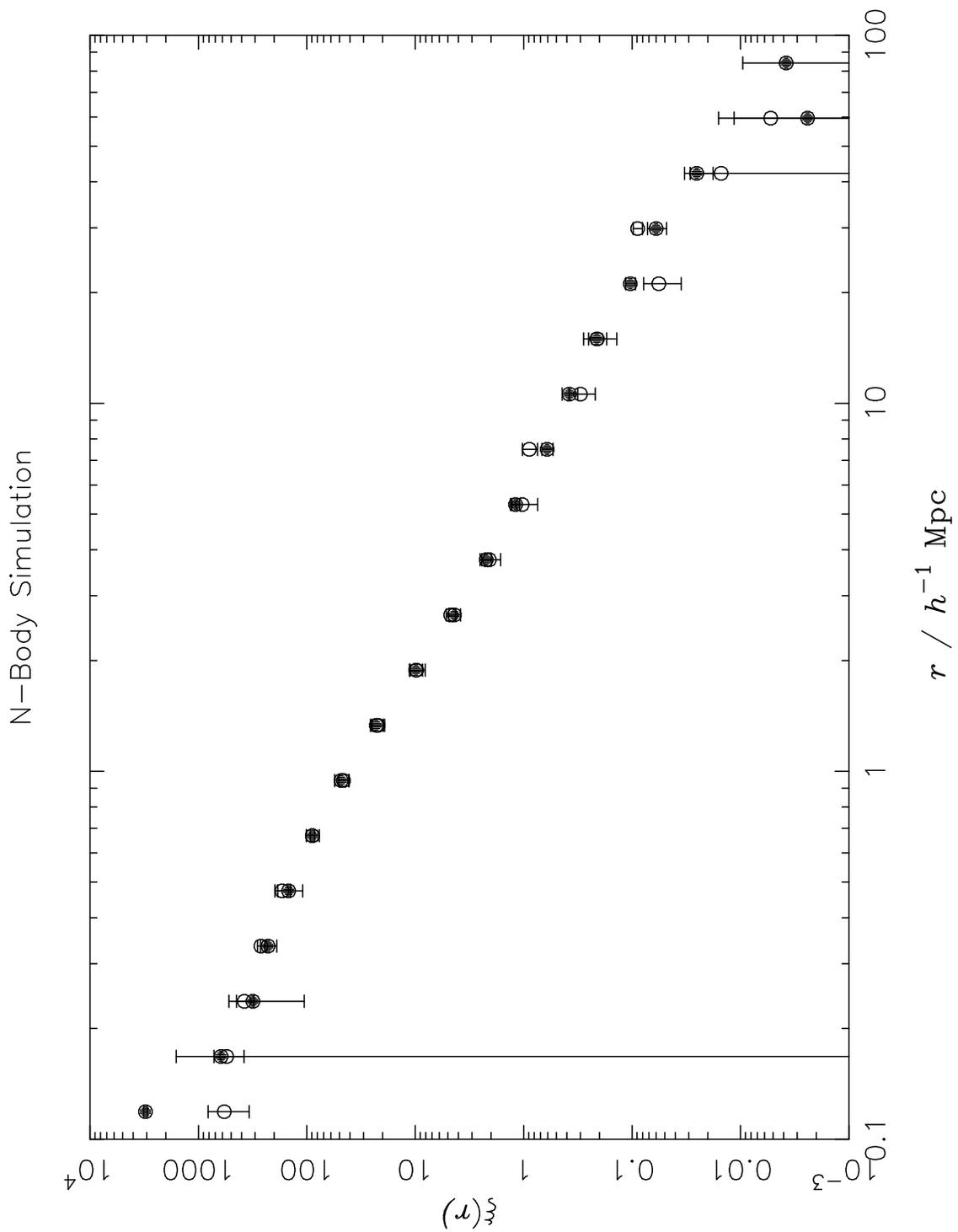

Figure 5:

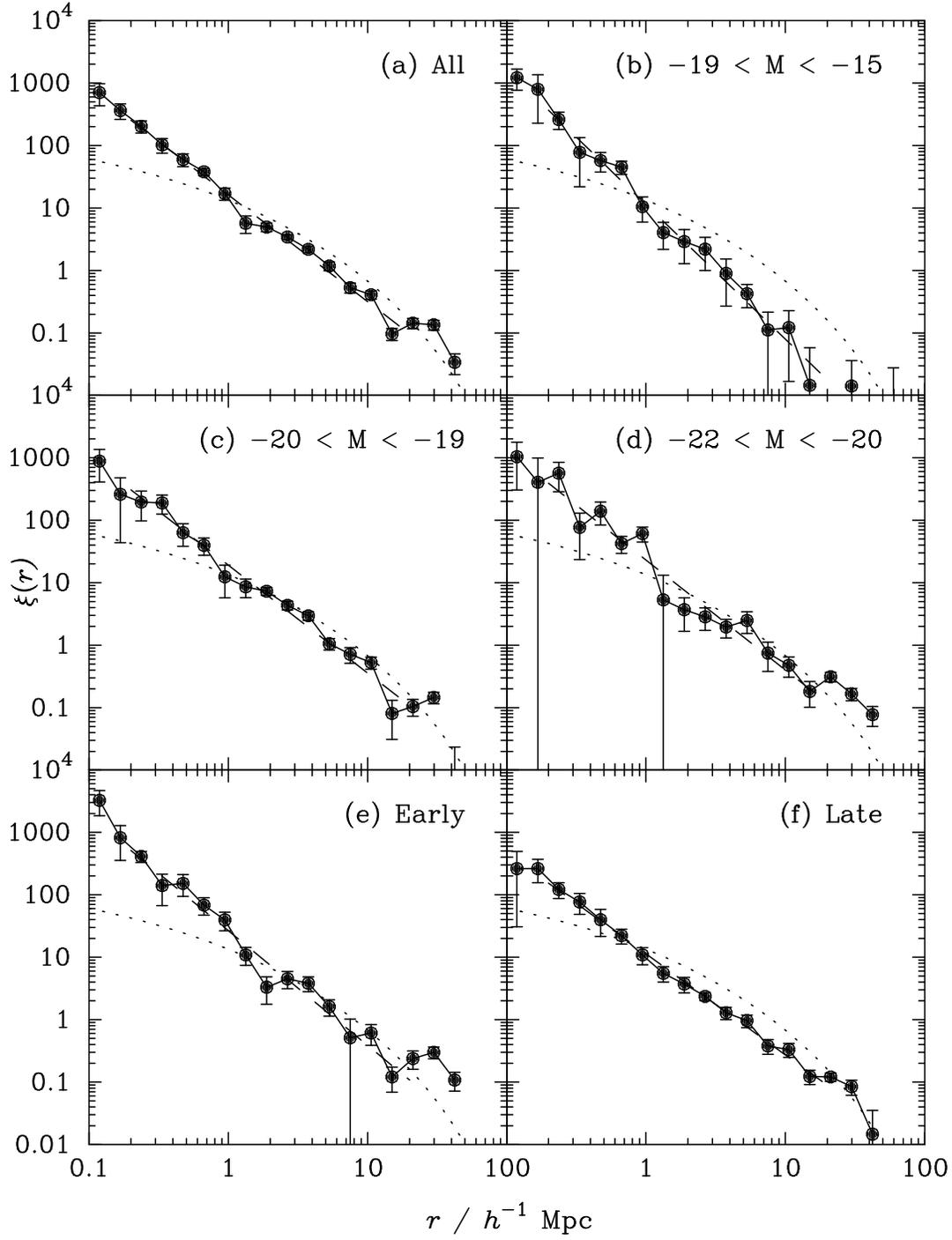

Figure 6:

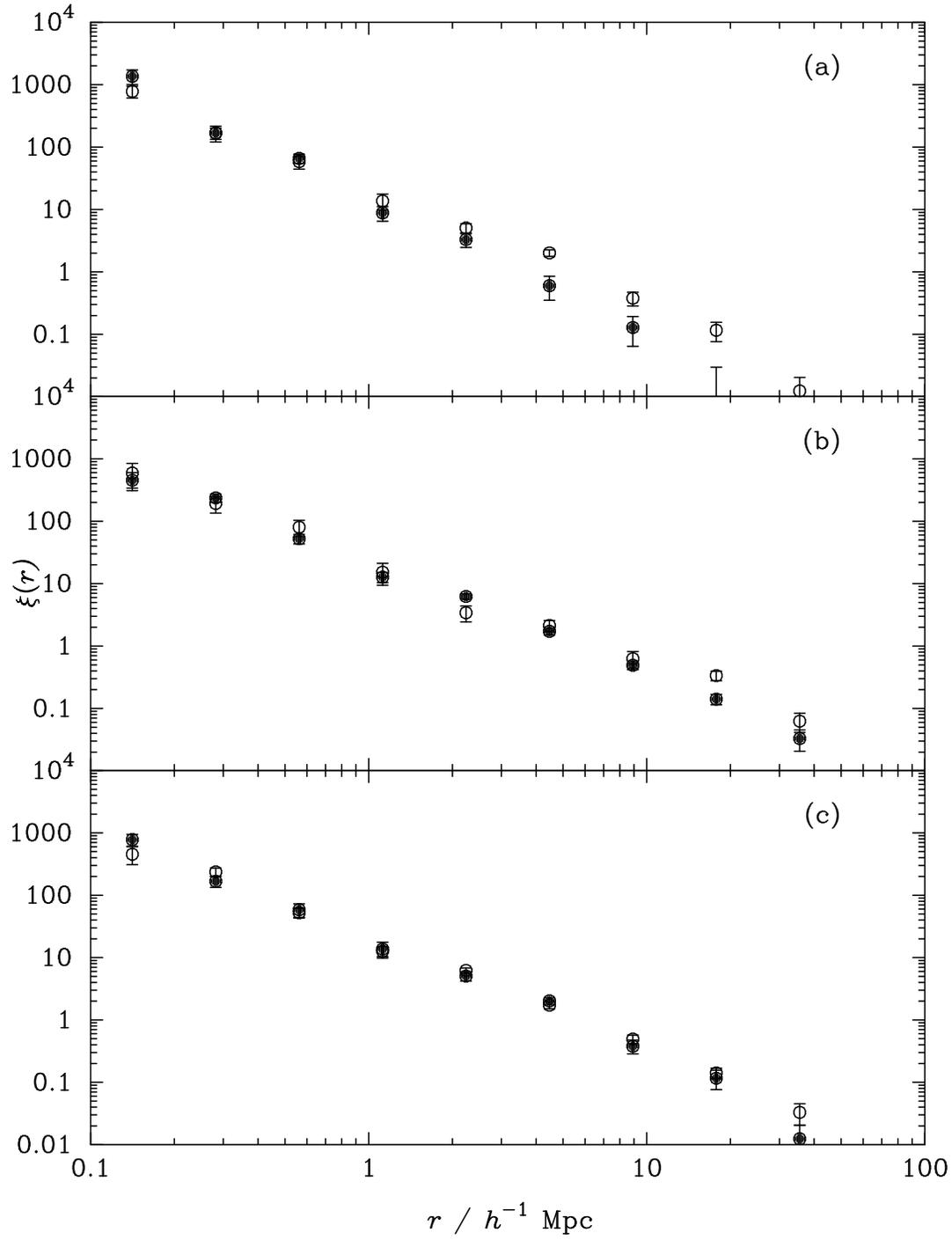

Figure 7:

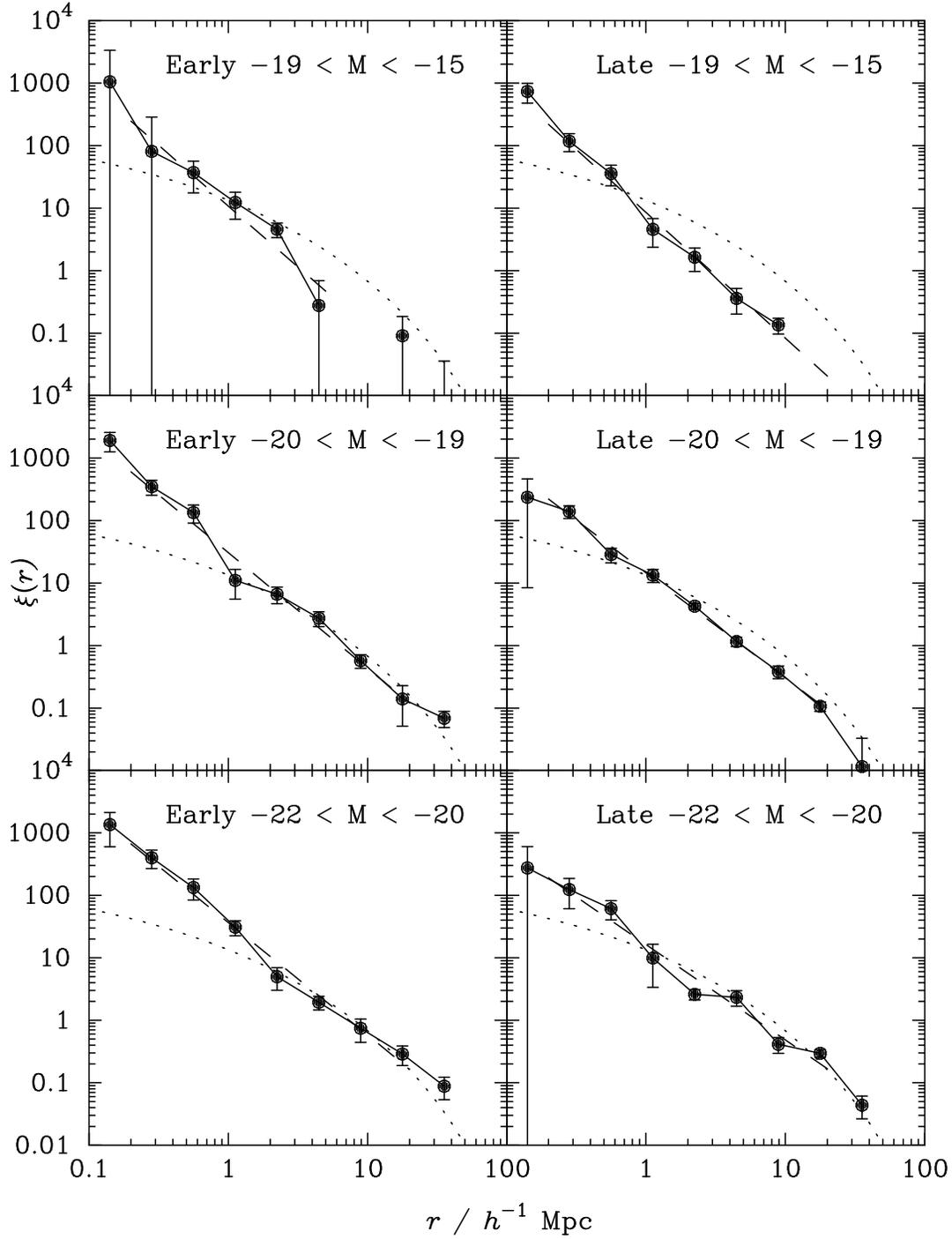

Figure 8:

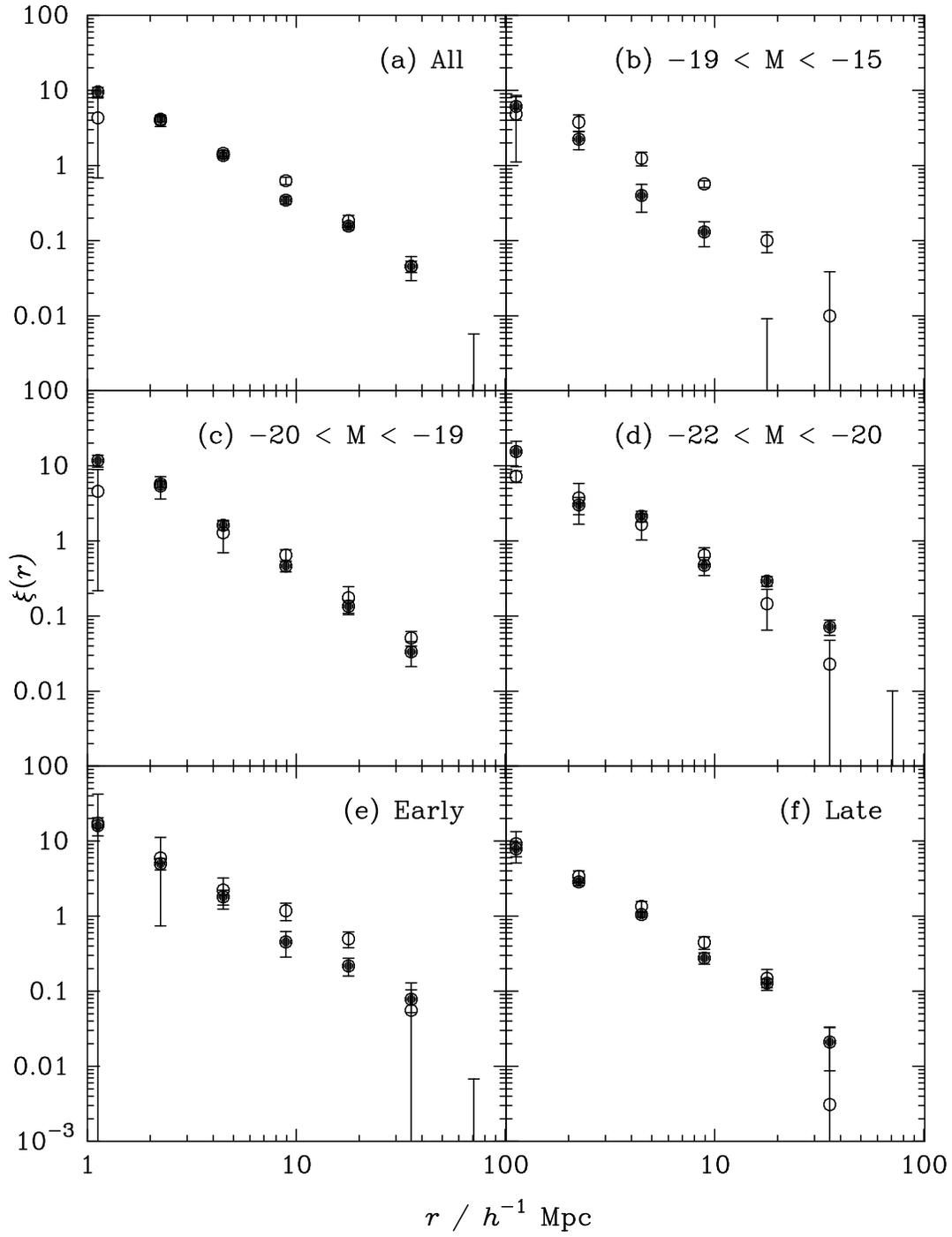

Figure 9: